**Cohesion and Shear Strength of Compacted Lunar and Martian Regolith Simulants.**

B. Dotson, D. Sanchez Valencia, C. Millwater, P. Easter, J. Long-Fox, D. Britt, and P. Metzger, University of Central Florida, Department of Physics, 4000 Central Florida Boulevard, Orlando, FL 32816; email: bdotson@knights.ucf.edu.

Shear strength and cohesion of granular materials are important geotechnical properties that play a crucial role in the stability and behavior of lunar and Martian regolith, as well as their terrestrial analog materials. To characterize and predict shear strength and cohesion for future space missions, it is also important to understand the effects of particle size distribution and density on these fundamental geotechnical properties. Generalized equations have been established using empirical data from direct shear measurements of lunar and Martian regolith simulants to quantify the effects of particle size distribution and density on cohesion and shear strength. Preliminary results are also presented highlighting the effects of atmospheric absorbed water on shear strength and cohesion when conducting experiments in atmospheric conditions on Earth. The results of this study show that cohesion increases exponentially with bulk density, while the exponential growth constant is also dependent on particle size distribution. The presence of absorbed atmospheric water can also produce non-monotonic effects on the shear strength of regolith, and it's influence on shear strength varies based on sample density for Earth-based experiments and applications. Together, these results highlight the importance of particle size distribution, bulk density, and atmospheric water content on the shear strength and cohesion of lunar and Martian simulants. This study also establishes generalized equations for predictive, testing, and modeling efforts for future space missions.

**1.0 Introduction**

As human and robotic space exploration missions strive to land heavier payloads on the surfaces of planetary bodies, build requisite exploration infrastructure, and leverage materials using *in situ* Resource Utilization (ISRU), understanding the shear strength and cohesion of regolith will be important. These geotechnical properties play a pivotal role in shaping surface activities and safety measures during planetary exploration, influencing mission design areas such as the construction of new rover wheels, landing pads, foundations for surface structures, traverse planning, and sampling considerations. Understanding the physical properties of lunar and Martian regolith will also be important for examining and mitigating the effects of plume surface interactions (PSI) during spacecraft landings.

Past missions provide valuable insights into the consequences of low cohesion and shear strength. The *Spirit* rover's entrapment on Mars in 2009, for example, was due to its wheels slipping in low-cohesion regolith and eventually resulted in mission termination (Matson, 2010; "NASA's Mars Rover Is Really Stuck," 2009; Wang et al., 2007). Similarly, the *Interior Exploration using Seismic Investigations, Geodesy and Heat Transport (InSight)* mission's *Heat Flow and Physical Properties Package (HP$^3$)* encountered difficulties burrowing into Martian regolith in 2019, primarily attributed to its low cohesion (Kedar et al., 2017; Spohn et al., 2022; Wippermann et al., 2020). Additionally, the Apollo 12 Lunar Module landed at the rim of Surveyor crater in 1969, causing damage to the Surveyor 3 spacecraft roughly 200m away, demonstrating how exceeding the shear strength and cohesion of lunar regolith when exposed to the exhaust from a landing rocket, can have detrimental effects (Clegg et al., 2014; Jaffe, 1971; J. Lane et al., 2012). These experiences underscore the critical need for accurate characterization and prediction of regolith's shear strength and cohesion for ensuring mission safety, optimizing resource utilization, and enabling sustainable human presence beyond Earth. Additionally, future exploration missions will benefit from refined PSI models based on regolith's shear strength and cohesion as functions of bulk density and particle size distribution.

1.1 Mohr-Coulomb Failure Criterion

Shear strength ($\sigma_s$) is a material property that represents the maximum, pre-failure, resistive force per area of a material to a shearing stress through constituent particle interactions. While dependent on the magnitude of normal stress applied ($\sigma_N$), shear strength is related to a material's cohesion (*c*) and internal friction angle ($\phi$) as described by the Mohr-Coulomb Failure Criterion in Equation 1 below (Labuz & Zang, 2012).

$$\sigma_S = \sigma_N \tan \phi + c \qquad (1)$$

As seen from this criterion, this model assumes a linear relationship between shear strength and normal stress that is also dependent on the material's properties, namely cohesion and internal friction angle. As such, the cohesion of a material is a measure of the inherent force holding particles together, without an applied normal stress and absent of contributions from internal friction. The shear strength and cohesion of a material, specifically the Mohr-Coulomb components of cohesion and angle of internal friction, are common inputs into engineering and computational models for construction, mission or instrumentation designs, or ISRU activities.

Other studies have examined the non-linear effects on cohesion in reduced gravity environments and without a normal stress to relate cohesion of a material to tensile strength, also a fundamental material property, and a non-linearity coefficient (K. Alshibli, 2017). Such effects may dominate in the lower gravity environments of the Moon or Mars, but additional research is needed. For the purposes of this study in terrestrial gravity and atmospheric conditions, these non-linear effects and material tensile strengths have not been considered.

Since cohesion is a measure of interparticle forces holding the constituents of a material together, these forces can be influenced by factors such as particle size, surface area in contact (which is related to shape and density), and electrostatic forces. For experiments on Earth, absorbed atmospheric water or moisture content will also impact the cohesive nature of a material, especially for granular materials like lunar and Martian regolith simulants (Fattah & Al-Lami, 2016; J. Long-Fox et al., 2023). As such, this study attempts to determine and quantify the effects of particle size distribution and bulk density on cohesion and shear strength using compacted lunar and Martian simulants, to help establish generalized equations for future design and computational models. To accurately apply experiments on Earth in atmospheric conditions to the lunar and Martian environments, this study also examines the effects of absorbed atmospheric water on shear strength and cohesion.

## 2.0 Methods

In order to determine the effects of particle size distribution and bulk density on the shear strength and cohesion of various granular materials for future space exploration efforts, the following sections highlight the techniques and procedures used on regolith simulants and glass bead controls in terrestrial atmosphere and gravity. These techniques and procedures were ultimately used to produce generalized equations for predicting the shear strength and cohesion of other samples, as long as the sample density and particle size distribution are known.

2.1 Regolith Simulants and Controls

Since it is not currently feasible to study *in situ* lunar and Martian regolith, and since only small amounts of regolith samples have been returned to Earth, laboratory simulants have been produced by the Exolith Laboratory at the University of Central Florida (UCF) as well as NASA's Kennedy Space Center (KSC) Swamp Works. Similarly, simulants have also been developed by the Simulant Development Lab (SDL) at NASA's Johnson Space Center (JSC), in collaboration with the United States Geological Service (USGS). These simulants mimic the particle size distribution, composition, and mechanical properties of actual lunar and Martian regolith from either returned samples or *in situ* measurements (McKay et al., 1993).

For Martian simulants, the following Exolith Simulants were examined: Mars Global Simulant (MGS-1), Jezero Simulant (JEZ-1), MGS-1 with hydrated clay (MGS-1C), and MGS-1 with polyhydrated sulfates (MGS-1S) (Cannon et al., 2019). Additional information on these Exolith Martian simulants can be found at https://sciences.ucf.edu/class/exolithlab, to include bulk chemical compositions and constituent sources. Given the similarity, composition, and particle sizes relative to robotic surface samples, these high fidelity simulates are expected to have similar geotechnical properties as those observed from previous Mars exploration missions (Landsman & Britt, 2020).

Simulant JSC-1A, from NASA's SDL, is comprised mainly of a plagioclase, pyroxene, and olivine mixture from mined volcanic ash and was used to simulate lunar regolith (K. A. Alshibli & Hasan, 2009; J. E. Lane et al., 2016). Black Point basaltic cinders (BP-1), consisting predominately of crushed balsalt, is another common lunar regolith simulant from NASA's KSC Swamp Works that was also examined in this study (Suescun-Florez et al., 2015). Similarly, the following lunar regolith simulants were used from UCF's Exolith Laboratory: Lunar Highlands Simulant (LHS-1), LHS-1 Dust Simulant (LHS-1D), Lunar Mare Simulant (LMS-1), LMS-1 Dust Simulant (LMS-1D), and LHS-1 with 25% anorthosite agglutinates by weight (LHS-1-25A) (Easter et al., 2022; J. Long-Fox et al., 2023; J. M. Long-Fox et al., 2023). Additional information on these Exolith lunar simulants can also be found at https://sciences.ucf.edu/class/exolithlab.

Of note, the LHS-1-25A simulant contains agglutinate anorthosite glass, produced by high temperature sintering with anorthosite and an iron matrix to mimic the composition, mineralogy, and mechanical properties associated with meteor impacts on the Moon. Since previous studies of lunar regolith contained roughly 15 to 60% agglutinate materials by weight, the examined LHS-1-25A sample likely serves as a rough approximation of lunar regolith in bulk composition (Papike et al., 1982). However, a more detailed study of LHS-1-25A agglutinate shapes and sizes, using a scanning electron microscope (SEM), is needed to fully understand the fidelity of LHS-1-25A as a true simulant for lunar regolith. Yet, as a newly produced simulant, this study examined geotechnical measurements of LHS-1-25A as an initial baseline.

Table 1. Planetary regolith simulants examined in this study.

| Moon | Mars | Custom |
|---|---|---|
| LHS-1 | MGS-1 | Glass Beads, 1mm |
| LHS-1D | JEZ-1 | Glass Beads, 40-80um |
| LHS-1D, Dried | MGS-1C | - |
| LHS-1-25A | MGS-1S | - |
| LMS-1 | - | - |
| LMS-1D | - | - |
| JSC-1A | - | - |
| BP-1 | - | - |

2.2 Particle Shapes

Particle shapes, including average aspect ratio (a), elongation (l), perimeter (p), sphericity (s), equivalent circular diameter ($x_{ECD}$), and convexity ($\psi_C$) were measured for each sample listed in Table 1 using a CILAS-1190 connected to a CETI Inverso TC 100, inverted optical microscope with an objective magnification of 40x and ExpertShape analysis software. This device uses an IDS uEye UT149xLE-C camera to digitally record microscope images with a numerical aperture of 0.10 and a resolution of 2.452 μm. For all images analyzed using the ExpertShape analysis software, a threshold value of 58% was used with a total particle number (N) greater than 700 particles of various sizes.

For the purposes of this study, the relative shapes of particle grains in regolith and custom simulants are cataloged only and not intentionally varied. It is expected that particle shape will have a lesser, but non-zero impact, on the overall cohesion and shear strength comparisons between lunar simulants and actual regolith on the Moon which is known to be more irregular and jagged in shape. However, additional research outside of this study is required to determine the overall effect of particle shape when comparing lunar regolith and lunar regolith simulants. Unlike the Moon, weathering processes on the Martian surface are expected to make Martian regolith simulants a suitable approximation for particle shapes, with similar geotechnical behavior.

2.3 Particle Size Distributions

Particle size distributions for the samples in Table 1 were measured using a Cilas-1190 laser-diffraction size analyzer with 0.50 g of each sample. The mass of each sample was measured using a BSM220.4 electronic balance with an advertised readability of 0.0001 g. Particle size distributions were measured for a total of 5 different samples (0.50 g each), selected randomly from containers of roughly 10-15 kg for each simulant, in order to find the average particle size distributions and associated standard deviations. It is important to note that the examined simulants do not contain course materials or particles above 2 mm in diameter, by design. While larger, coarse materials may impact the overall cohesion and shear strength of the material, this study focuses purely on the granular material in lunar and Martian regolith.

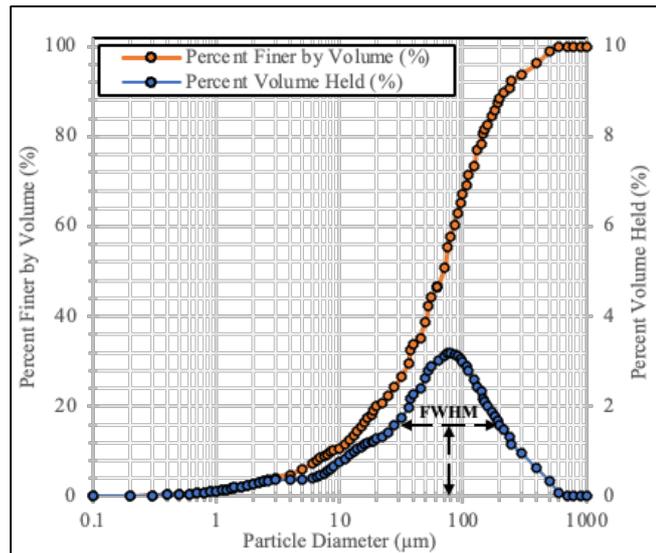
*Figure 1. Example cumulative and differential particle size distributions for LHS-1 with FWHM.*

Using the differential particle size distribution or percent volume held, rather than the cumulative particle size distribution or percent finer by volume as commonly reported, a measure of the Full Width at Half Maximum for the specific particle size distribution (FWHM) was obtained for each sample using a 3-5 term, Gaussian, least squares fit in MATLAB, as shown in Figure 1 above. While traditional characterizations of particle size distributions leverage curvature coefficients, the FWHM is used in this study as an attempt to characterize the material's particle size distribution with a single number. The average particle size curvature coefficients were also recorded and cataloged using the CETI Inverso TC 100 as well.

2.4 Direct Shear Measurements

Direct shear measurements were conducted using the simulants and controls listed in Table 1, in accordance with the American Society for Testing and Materials (ASTM) D3080: Standard Test Method for Direct Shear Test of Soils Under Consolidated Drained Conditions (ASTM International, 2011). These direct shear measurements were conducted in a custom-built polycarbonate direct shear box as shown in Figure 2 below. This apparatus was used instead of commercial-off-the-shelf (COTS), geotechnical lab equipment, as the desired range of normal stress was only achievable with such a custom-built setup. This apparatus uses two halves of a vertically stacked shear box, with interior shearing surface dimensions of 10.17 cm by 10.17 cm. Since the shear strength of a sample is a material property, the size of shearing area is not expected to impact the overall shear measurement from this custom-built apparatus.

The bottom half of the direct shear box was attached to a rail system, with an HP-500 force gauge (advertised resolution of 0.01 N) fixed to an Actuonix L16-R linear translation servo motor that was controlled by an Arduino UNO R3 microcontroller in order to shear the boxes at a constant speed. For each test, a normal stress was applied by using calibrated sheets of aluminum in a separate polycarbonate box, gently placed onto the simulant or control sample. At least 4 different normal stresses ranging from 0.097 to 0.68 kPa were examined in this manner, with roughly 5 measurements per normal stress per sample density.

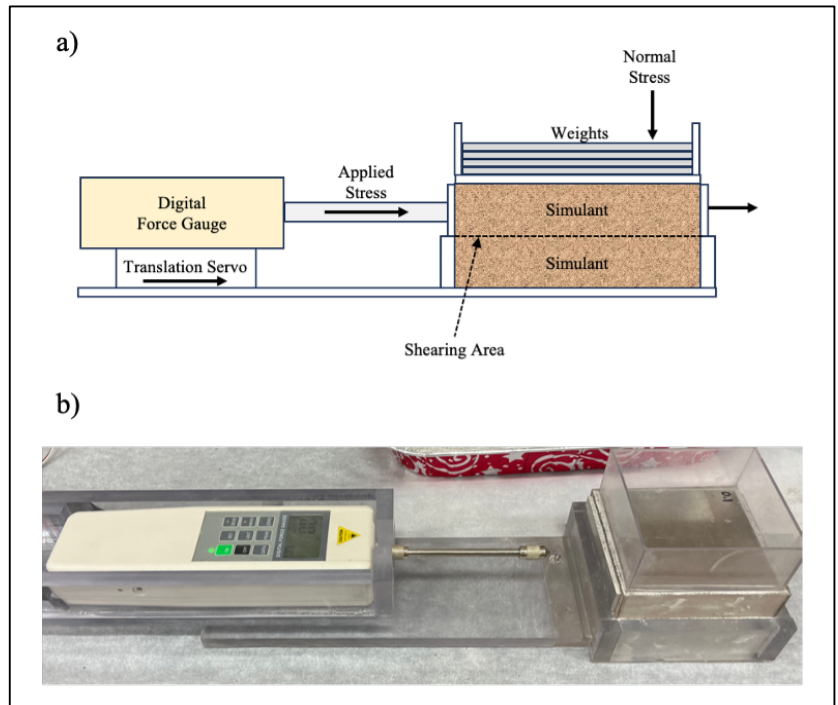

*Figure 2. a) Design of direct shear apparatus and b) experimental setup with an HP-500 force gauge, translation servo, and calibrated normal stress of 0.1 kPa placed on top of LHS-1 simulant. The force gauge measures the maximum force just prior to material failure, as the two halves of the sample box undergo a shearing motion.*

      In order to control the sample density during the direct shear tests, a known mass of simulant was sprinkled into the box using a flat, plastic scoop, in an effort to achieve uncompressed, average bulk density. The density of the sample was recorded by weighing the simulant in the shear box (of known volume) on a Vevor electronic scale with an advertised readability of 0.001 g. If the sample density was below the target density, mechanical vibration using a vibration table at 20-40 Hz was used to slightly compact the material in order to add additional mass into the known volume of the shear box. If the sample density was inadvertently greater than the target density, the entire sample box would be emptied and refilled with fresh sample. The duration of the mechanical vibration was varied, depending on the target sample density. For slight compaction, roughly 0.5 to 1 second pulses at 20 Hz were used. Conversely, extended vibration durations, on the order of 3-5 minutes, at 20-40 Hz were used in conjunction with a normal stress of 0.68 kPa (the same aluminum sheets described above) for higher target densities. Aside from this 0.68 kPa normal stress, excessive normal stresses or pushing on the sample were not used due to the desire to achieve a uniform density distribution throughout the sample shear box. No additional steps were taken to measure or control sample density as a function of depth within the shear box, but since the depth of the custom-built apparatus was 6 cm, the density was assumed to be uniform throughout.

2.5 Absorbed Atmospheric Water

      All samples were exposed to atmosphere in a climate-controlled laboratory environment during measurements, with an average room temperature of 22 °C and no direct control or measurement of relative humidity. Between measurements, the samples were stored in sealed

containers without prolonged exposure to the laboratory environment. In order to mitigate the effects of humidity changes, direct shear measurements for an individual density were taken for each sample at different normal stresses generally on the same day, if possible. Given the climate-control of the laboratory, it is believed that humidity changes and amounts of absorbed water were relatively constant throughout the study.

To highlight the impacts of absorbed atmospheric water on the shear strength and cohesion of materials, a separate and limited series of direct shear measurements were conducted using only LHS-1D (a very hydroscopic material) with and without atmospheric water absorbed. Direct shear measurements were taken at different densities, as described in Section 2.4 above, while exposing the samples to the laboratory environment. These measurements were then repeated using the same LHS-1D samples, after baking the simulant for 24 hours at 140°C, with the simulant still hot and transferred immediately from the oven to the shear box. In a similar fashion, the LHS-1D was weighed prior to and immediately after baking in order to determine the percent of absorbed atmospheric water. Being a limited initial investigation, no steps were taken to control or measure errors associated with buoyant plume effects when weighing hot samples, which could be a potential source of error in the mass measurements. Such effects are anticipated to be considered in more detail with future studies, an efforts are already underway to repeat similar direct shear measurements in a vacuum chamber. However, since the amount of absorbed atmospheric water did not vary throughout the limited LHS-1D tests, this highlights the consistency of relative humidity over the course of an individual sample's direct shear measurements. This same process was repeated for different densities of LHS-1D as well.

**3.0 Results**

In order to understand the relationships between geotechnical measurements and physical properties of lunar and Martian simulants, as well as representative controls, measured values for particle shapes, size distributions, cohesion, and shear strength are presented in the following sections. Confidence intervals reported throughout this study represent one standard deviation.

3.1 Particle Shapes

Particle shape parameters for each of the examined granular materials are presented in Table 2 below. From these measured values, the aspect ratio for all examined samples was measured between 0.647 and 0.960, with an average aspect ratio of $0.614 \pm 0.084$. Of all of the examined samples, the highest aspect ratio was associated with 1mm glass beads, indicating these specific particles were mostly spherical in shape. For comparison with returned lunar samples, Heywood (1971) reported aspect ratios between 0.725 and 0.855 for Apollo 12 samples, while Tsuchiyama et al. (2022) measured aspect ratios between 0.551 and 0.744 for both Apollo and Luna samples (Rickman et al., 2012; Tsuchiyama et al., 2022).

As shown in Table 2, the elongation for each examined sample ranged from 0.040 and 0.353, with an average elongation of $0.309 \pm 0.083$. The measured elongation for 1mm glass beads was the smallest of all samples, again suggesting a particle shape close to a perfect sphere. While not directly reported, using the measured aspect ratios from Heywood (1971) and Tsuchiyama et al. (2022) reveals a calculated value of elongation between 0.145 – 0.275 and 0.256 - 0.449, respectively (Rickman et al., 2012; Tsuchiyama et al., 2022). The average sphericity for all samples is $0.594 \pm 0.088$, with 1mm glass beads having a shape closest to spherical (s=1).

While values for convexity, perimeter, and equivalent circular diameter have not been specifically reported for actual lunar and Martian regolith, these measurements for lunar and Martian simulants are captured in Table 2 below, including glass bead controls as a comparison. From these measured values, the examined simulants generally have a more-spherical shape, without excessive convexity or irregular curvatures. For dusty simulants with smaller particle sizes, a similar reduction in perimeter and equivalent circular diameter is also noted. From these measured results, the perimeter, convexity, and equivalent circular diameter for non-dusty, Exolith simulants of LHS-1 and LMS-1 are generally comparable to JSC-1A and BP-1 lunar simulants produced by NASA. Moreover, likely due to similar manufacturing techniques, Martian simulants produced by Exolith lab also have a similar perimeter and equivalent circular diameter when compared to dusty regolith simulants.

Table 2. Measured particle shapes, including: average aspect ratio (a), elongation (l), sphericity (s), convexity ($\psi_C$), perimeter (p), and equivalent circular diameter ($x_{ECD}$).

| Type | Simulant | a | l | s | $\psi_C$ | p, μm | $x_{ECD}$, μm |
|---|---|---|---|---|---|---|---|
| Moon | LHS-1 | 0.655 | 0.345 | 0.561 | 2.089 | 128.51 | 25.63 |
| Moon | LHS-1D | 0.684 | 0.316 | 0.657 | 0.814 | 52.18 | 13.09 |
| Moon | LHS-1-25A | 0.665 | 0.335 | 0.608 | 0.997 | 72.01 | 16.37 |
| Moon | LMS-1 | 0.652 | 0.348 | 0.554 | 1.088 | 90.92 | 21.21 |
| Moon | LMS-1D | 0.669 | 0.331 | 0.591 | 0.764 | 75.95 | 18.16 |
| Moon | JSC-1A | 0.655 | 0.345 | 0.545 | 1.705 | 87.53 | 19.10 |
| Moon | BP-1 | 0.654 | 0.346 | 0.528 | 1.304 | 134.31 | 27.56 |
| Mars | JEZ-1 | 0.667 | 0.335 | 0.592 | 1.410 | 134.48 | 27.73 |
| Mars | MGS-1 | 0.647 | 0.353 | 0.538 | 1.462 | 122.47 | 23.40 |
| Mars | MGS-1S | 0.665 | 0.335 | 0.585 | 0.941 | 79.68 | 17.96 |
| Mars | MGS-1C | 0.674 | 0.326 | 0.582 | 0.934 | 78.72 | 18.08 |
| Control | Glass Beads, 1mm | 0.960 | 0.040 | 0.858 | 0.981 | 3,607.17 | $1.035 \times 10^6$ |
| Control | Glass Beads, 44-80μm | 0.743 | 0.257 | 0.521 | 0.838 | 212.93 | 48.37 |

3.2 Particle Size Distributions

Measuring particle size distributions for each examined sample, the average particle size ($X_{ave}$), FWHM and corresponding differential particle size, Gaussian peak location ($X_{FHWM}$), as well as curvature coefficients are reported in Table 3. Of particular note, the FWHM and $X_{FHWM}$ for each sample differ from the measured average particle size and 50% curvature coefficient ($D_{50}$). On average, the $D_{50}$ curvature coefficient is greater than $X_{ave}$ by a factor of roughly 2.65, and greater than $X_{FHWM}$ by a factor of roughly 5. However, as expected, the smaller grained samples (LHS-1D and LMS-1D) have smaller values for $X_{ave}$, $X_{FHWM}$, and $D_{50}$ when compared to larger grained materials of the same sample (LHS-1 and LMS-1). Also of note, there is a large range of 10% and 90% curvature coefficients ($D_{10}$ and $D_{90}$, respectively), where the particle sizes captured between $D_{10}$ and $D_{90}$ are roughly twice that of FWHM. Taken together, the values presented in Table 3 represent multiple ways of describing and comparing particle size distributions for a number of different samples including glass bead controls, as well as lunar and Martian simulants.

Table 3. Particle Size Distributions.

| Type | Simulant | $X_{ave}$, μm | $X_{FWHM}$, μm | FWHM, μm | $D_{10}$, μm | $D_{50}$, μm | $D_{90}$, μm |
|---|---|---|---|---|---|---|---|
| Moon | LHS-1 | 92 | 71.93 | 151.27 ± 1.05 | 124.73 | 293.48 | 405.91 |
| Moon | LHS-1D | 30 | 12.41 | 18.63 ± 0.03 | 18.27 | 32.96 | 37.46 |
| Moon | LHS-1-25A | 65 | 73.77 | 207.30 ± 2.04 | 58.84 | 141.14 | 275.71 |
| Moon | LMS-1 | 90 | 77.43 | 175.03 ± 1.07 | 53.91 | 116.50 | 277.21 |
| Moon | LMS-1D | 35 | 10.99 | 200.67 ± 1.89 | 84.53 | 211.50 | 428.69 |
| Moon | JSC-1A | 164 | 55.59 | 280.08 ± 0.25 | 57.06 | 225.94 | 343.05 |
| Moon | BP-1 | 198 | 51.88 | 150.22 ± 0.20 | 92.10 | 228.91 | 319.80 |
| Mars | JEZ-1 | 68 | 70.79 | 99.55 ± 4.99 | 120.30 | 269.56 | 512.48 |
| Mars | MGS-1 | 90 | 48.19 | 72.92 ± 9.23 | 109.06 | 232.36 | 384.08 |
| Mars | MGS-1S | 119 | 69.02 | 242.41 ± 10.5 | 72.05 | 145.09 | 213.36 |
| Mars | MGS-1C | 25 | 12.76 | 86.63 ± 1.11 | 73.83 | 190.86 | 342.97 |
| Control | Glass Beads, 1mm | 1,008 | 1,027.30 | 6.90 ± 0.96 | 1,140.54 | 1,148.20 | 1,156.15 |
| Control | Glass Beads, 44-80μm | 53 | 57.49 | 37.64 ± 0.20 | 53.13 | 90.25 | 110.44 |

3.3 Cohesion of Lunar and Martian Simulants

Using the methods described above, direct shear measurements were taken for the samples listed in Table 1 with different normal stresses and densities, in order to determine sample cohesion as a function of density. An example of direct shear measurements taken for LHS-1 at a density of 1.60 g/cm$^3$ is shown in Figure 3. In this example, the maximum shear strength before failure varies linearly with normal stress (normal load force per area) as predicted by the Mohr- Coulomb failure criteria in Equation 1 above. For this specific density, the cohesion of LHS-1 is calculated to be 535.31± 11.60 Pa by assuming a linear relationship, in order to determine the shear strength with no normal stress applied. A similar calculation was repeated for different densities and samples in order to produce Figures 4 and 5 below, showing the cohesion of lunar simulants, as well as Martian simulants and control samples, as a function of bulk density.

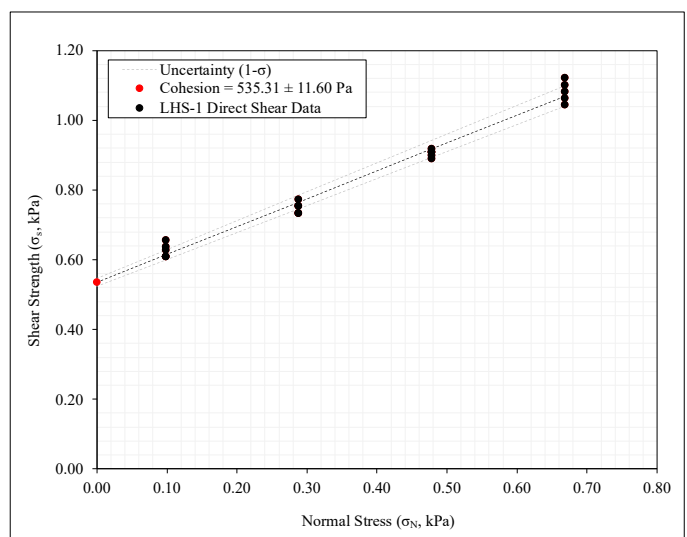

*Figure 3. Direct shear results for LHS-1 at 1.60 g/cm$^3$.*

As shown with examined lunar simulants in Figure 4, cohesion increases exponentially with an increase in sample density as described by Equation 2 below, where $\rho$ is sample density (including porosity effects), $A_1$ is a constant, and k is the exponential growth factor.

$$C(\rho) = A_1 e^{k*\rho} \quad (2)$$

For dusty simulants LHS-1D and LMS-1D, the measured densities range from roughly 0.80 - 1.30 g/cm³ while cohesion varied between roughly 0.116 – 1.289 kPa at different densities. For larger particle lunar simulants, measured densities ranged from roughly 1.32 – 1.97 g/cm³ while cohesion varied between roughly 0.185 – 1.672 kPa at different densities. Of the examined lunar simulants, LHS-1D had the lowest uncompressed bulk density, while LMS-1 had the highest uncompressed density and highest observed cohesion. Previously published cohesion values for LHS-1, LHS-1D, LMS-1, and LMS-1D by Millwater et al., 2022 are also shown in Figure 4, and agree with the measured results from this study.

As shown in Figure 4, a change in cohesion of roughly 15% was observed between dried LHS-1D and the same sample with ~0.4% absorbed atmospheric water by mass. These results are discussed in more detail in Section 4.4 below. When comparing cohesion as a function of density for LHS-1 and LHS-1-25A (LHS-1 with 25% agglutinates), no significant difference in cohesion was noted for lower relative densities. Maximum cohesion decreased by roughly 10% for LHS-1-25A at higher relative densities when compared to LHS-1, with the maximum observed density of 1.80 and 1.75 for LHS-1 and LHS-1-25A, respectively. This further demonstrates that adding 25% agglutinates to LHS-1 reduced the achievable range of bulk densities, and thus, the achievable range of sample cohesion. However, LHS-1-25A had a higher cohesion at any given density compared to LHS-1, as shown in Figure 4, likely due to higher interlocking of agglutinates and particles.

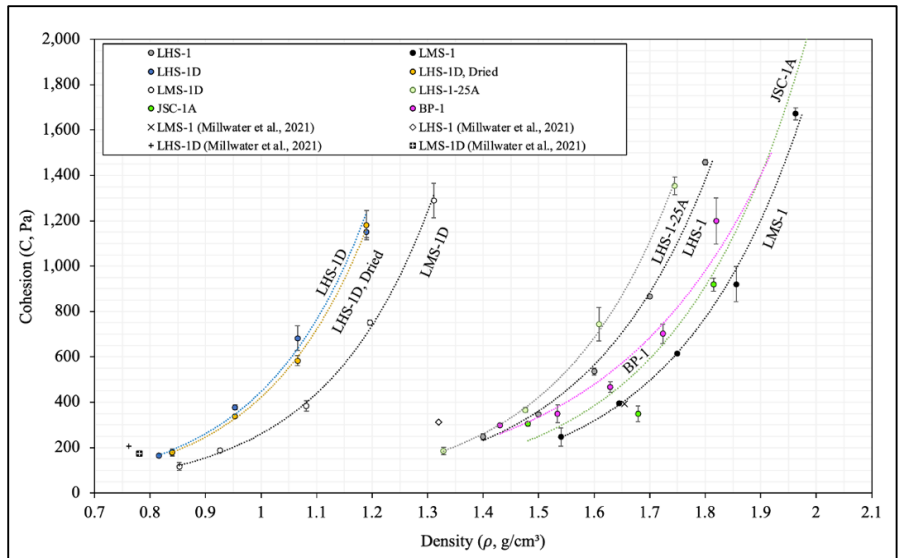

*Figure 4. Cohesion of lunar simulants as a function of density with exponential fits using Equation 2.*

As shown with examined Martian simulants and controls samples in Figure 5, cohesion increases exponentially with an increase in sample density. For Martian simulants, the measured densities range from roughly 1.03 - 1.86 g/cm³ while cohesion varied between roughly 0.179 – 1.406 kPa at different densities. Despite having different average particle sizes by as much as 24% (FWHM), the exponential fit for cohesion as a function of density for MGS-1 and MGS-1S samples was roughly the same, as shown in Figure 4. However, based on observed variance in the measured values for shear strength as a function of normal stress, higher relative densities for MGS-1S did result in larger uncertainty for cohesion as depicted. For both the 1mm and 44-80 µm glass beads, the achievable range of relative densities were less than that of Martian or lunar simulants, given a more uniform particle size distribution, with measured densities between roughly 1.40 – 1.60 g/cm³. This decrease in achievable relative densities also coincided with a smaller range of observed cohesion values, varying with density between 0.181 – 0.349 kPa for both sizes of glass beads.

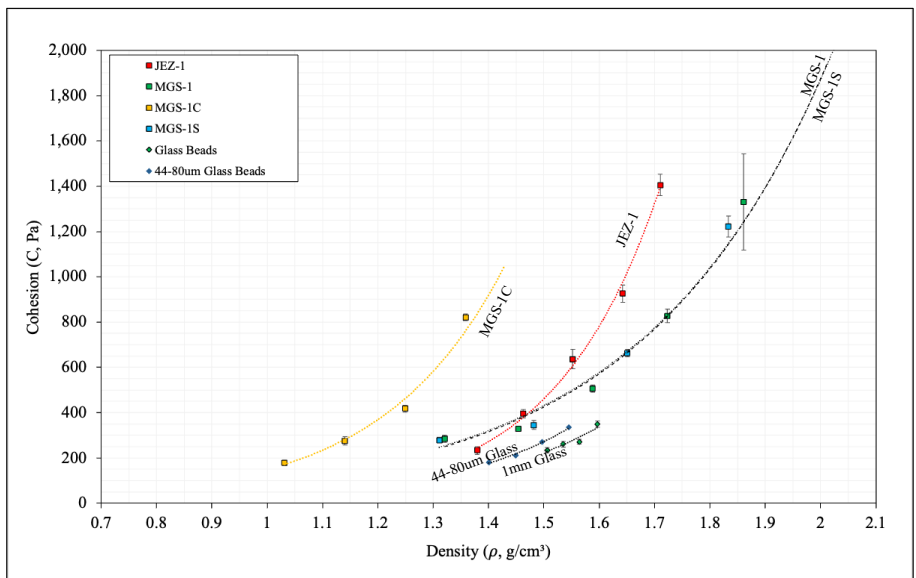

*Figure 5. Cohesion of Martian simulants and control samples as a function of density with exponential fits using Equation 2.*

A summary of the exponential fit parameters for all examined materials, as well as the associated error and coefficient of determination ($R^2$), are reported in Table 4 below. These fit parameters are derived from the empirical data in Figures 4 and 5, and collectively demonstrate the exponential increase in cohesion as a function of density as described by Equation 2. The uncompressed bulk density ($\rho_b$) and compressed bulk density ($\rho_{b,max}$) for each sample is also reported in Table 4, where $\rho_{b,max}$ is the highest density examined during this study. As shown, the individual exponential fits of empirical data generally have $R^2$ values greater than 0.98 for most samples, with the exception of the exponential fit for 1mm glass bead data yielding an $R^2$ value of 0.9095. These empirical fit parameters were then leveraged to develop a generalized model for cohesion as a function of density, FHWM, and $\rho_b$, as discussed in Section 3.5.

Table 4. Exponential fit parameters for cohesion as a function of density.

| Type | Simulant | $A_1$ | k | $R^2$ | $\rho_b$, g/cm$^3$ | $\rho_{b,max}$, g/cm$^3$ |
|---|---|---|---|---|---|---|
| Moon | LHS-1 | 0.45 ± 0.36 | 4.46 ± 0.40 | 0.9914 | 1.32 | 1.80 |
| Moon | LHS-1D | 2.12 ± 1.34 | 5.35 ± 0.33 | 0.9833 | 0.82 | 1.19 |
| Moon | LHS-1D, Dried | 2.00 ± 0.73 | 5.35 ± 0.32 | 0.9991 | 0.84 | 1.19 |
| Moon | LHS-1-25A | 0.31 ± 0.14 | 4.82 ± 0.22 | 0.9989 | 1.33 | 1.75 |
| Moon | LMS-1 | 0.27 ± 0.09 | 4.42 ± 0.31 | 0.9854 | 1.54 | 1.96 |
| Moon | LMS-1D | 1.34 ± 0.40 | 5.20 ± 0.19 | 0.9972 | 0.85 | 1.31 |
| Moon | JSC-1A | 0.39 ± 0.04 | 4.31 ± 0.87 | 0.9587 | 1.48 | 1.98 |
| Moon | BP-1 | 1.57 ± 0.44 | 3.57 ± 0.57 | 0.9666 | 1.43 | 1.82 |
| Mars | JEZ-1 | 0.17 ± 1.34 | 5.27 ± 0.33 | 0.9833 | 0.82 | 1.71 |
| Mars | MGS-1 | 4.96 ± 0.34 | 2.97 ± 0.22 | 0.9989 | 1.32 | 1.86 |
| Mars | MGS-1S | 5.28 ± 1.14 | 2.93 ± 0.22 | 0.9854 | 1.31 | 1.83 |
| Mars | MGS-1C | 1.56 ± 0.66 | 4.56 ± 0.66 | 0.9863 | 1.03 | 1.36 |
| Control | Glass Beads, 1mm | 0.48 ± 0.62 | 4.10 ± 1.23 | 0.9095 | 1.51 | 1.60 |
| Control | Glass Beads, 44-80μm | 0.39 ± 0.17 | 4.35 ± 0.33 | 0.9924 | 1.40 | 1.55 |

3.4 Internal Friction Angle

The calculated internal friction angle for all examined granular samples is presented in Figure 6, using direct shear measurements at various densities. Previously published results from Millwater et al. (2021) are also presented in Figure 6 and are comparable to internal friction angle results from this study. Calculated values for internal friction angle ranged from roughly 15 - 65 degrees for all samples considered. Based on observed variance in the measured values for shear strength as a function of normal stress, higher relative densities for MGS-1 did result in larger uncertainties for internal friction angle with densities near 1.86 g/cm$^3$, as depicted.

Internal friction angle varied differently for certain samples as a function of sample density. For example, with an increase in sample density, the calculated internal friction angle for MGS-1C increased slightly to roughly 29 degrees at 1.14 g/cm$^3$, before decreasing to roughly 25 degrees at 1.35 g/cm$^3$, as depicted. Conversely, the calculated internal friction angle for MGS-1S gradually increased from roughly 34 degrees at 1.31 g/cm$^3$ to near 53 degrees at 1.83 g/cm$^3$, as shown. Collectively, this implies that not all samples have the same relationship between internal friction angle and density, likely owing to the particle size distributions and the amount of fine-grained particles in each sample as shown by Easter et al. (2022).

In an attempt to generalize the relationship between internal friction angle and density, a second-order polynomial was fit to this empirical data with an $R^2$ value of 0.652. This fit model is shown in Figure 6, and Equation 3 below, including one standard deviation errors for each fit parameter. While this empirical fit generally characterizes a broad relationship between internal friction angle and sample density, this fit is limited to sample densities between 0.70 – 2.10 g/cm$^3$ as shown in Figure 6.

$$\varphi(\rho) = (20.86 \pm 8.07)\rho^2 + (-33.932 \pm 21.96)\rho + (38.93 \pm 14.48) \quad (3)$$

From this empirical fit, Equation 3 above can be used to determine the internal friction angle of a general sample, as a function of sample density. Of particular note, sample density in this instance is bulk density which inherently includes effects from porosity.

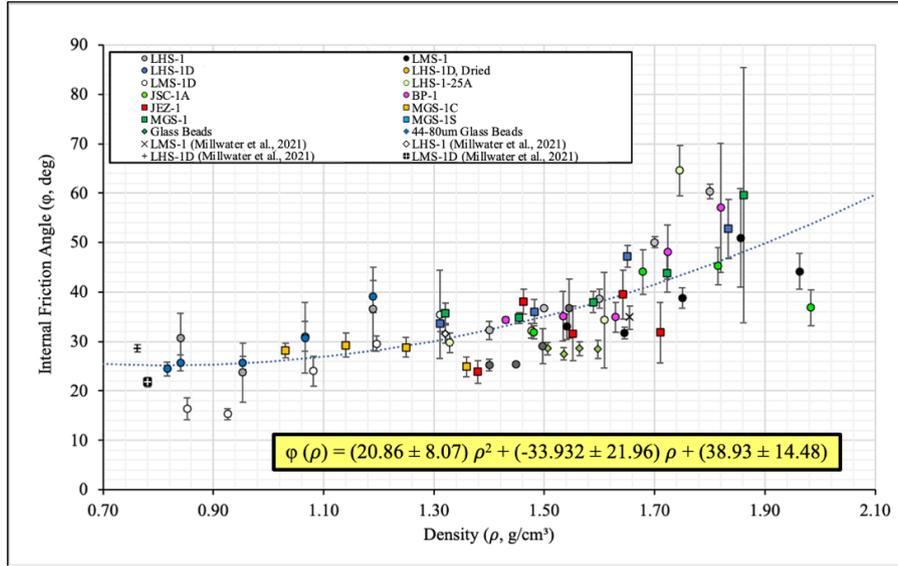

*Figure 6. Internal friction angle of simulants and control samples as a function of density with polynomial fit in Equation 3.*

3.5 Generalized Equation for Cohesion and Shear Strength

To establish a generalized equation for shear strength, exponential fit parameters from empirical data of all samples as listed in Tables 3 and 4 were compared to uncompressed bulk density and FWHM for all of the examined granular samples in Table 1. More specifically, the exponential fit parameters k and $A_1$ are shown in Figure 7 as a function of FWHM and $\rho_b$, respectively. From this empirical data, the exponential growth constant, k, increases linearly with an increase in FWHM as described in Equation 4 below. The exponential fit constant, $A_1$, decreases exponentially as a function of uncompressed bulk density as shown in Equation 5 below. In both Equations 4 and 5, the confidence intervals used for fit parameters represents one standard deviation.

$$k\ (FHWM) = (0.0012 \pm 0.0025)\ FWHM + (4.42 \pm 0.40) \quad (4)$$

$$A_1\ (\rho_b) = (70.37 \pm 7.25)\ e^{(-3.81 \pm 0.12)\ \rho_b} \quad (5)$$

Of note, the $R^2$ values of the empirical curve fits that produced Equations 4 and 5 are 0.225 and 0.328, respectively. These lower $R^2$ values are likely a result of the large scatter in the empirical data which is shown in Figure 7. Both constant $A_1$ and the exponential growth constants for MGS-1S and MGS-1 appear to deviate from trend lines in Figures 7A and 7B, likely due to the inconsistency of particle size distributions when randomly sampled. In performing measurements of particle size distributions for these samples, the differential and cumulative particle size distributions demonstrated a larger variance when examined across random samples. For consistency, the data points for MGS-1S and MGS-1 have been included in the empirical fits producing Equations 4 and 5, contributing to the $R^2$ values reported above. Additional discussion on this error is provided in Section 4.5 below.

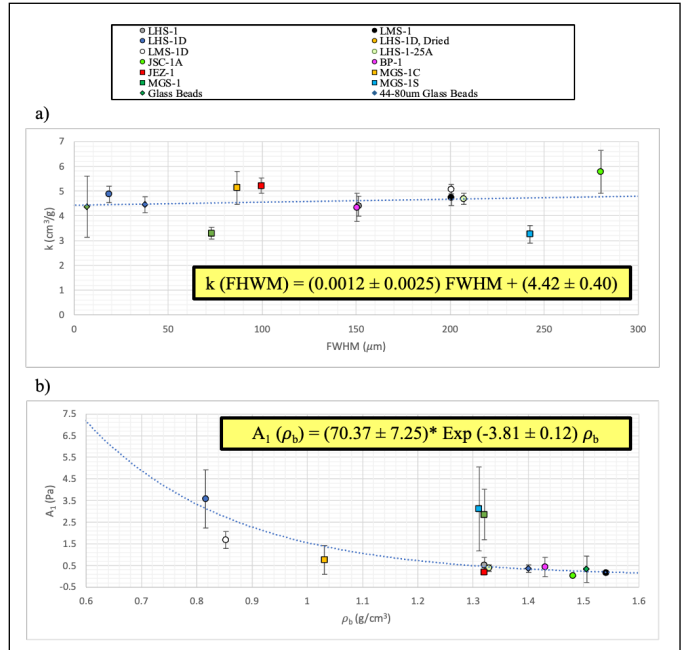

*Figure 7. Exponential fit parameters for simulants and control samples as a function of a) Full-Width-at-Half-Max (FWHM) for particle size distribution, and b) uncompressed bulk density ($\rho_b$).*

By substituting Equations 4 and 5 into Equation 2, a generalized equation for cohesion as a function of FWHM, $\rho_b$, and $\rho$ is produced. Such a relationship provides a means of estimating the cohesion of granular samples as a function of density where the FWHM of the particle size distribution and uncompressed bulk density are known. Similarly, combining Equations 1-5 produces a generalized equation for shear strength of a granular material as a function of normal stress and bulk density (including porosity), shown by Equation 6, where the FWHM of the particle size distribution and uncompressed bulk density are known. These relationships also allow for an examination of the sensitivities of cohesion and shear strength of granular materials to variations in bulk density, particle size distribution (FWHM), and normal stress as described in Sections 4.1-2 below.

$$\sigma_S (\sigma_N, \rho, \rho_b, \text{FWHM}) = f = \sigma_N \tan[\varphi(\rho)] + c(\rho, \rho_b, \text{FWHM}) \qquad (6)$$

As an example, by substituting measured values for FWHM and $\rho_b$ of LHS-1 into Equation 6, a 3-dimenstional representation of shear strength as a function of normal stress and density are shown in Figure 8, with direct shear measurements overlaid. From Figure 8, it is clear that applying Equation 6 produces a fit that closely matches the empirical data measured during direct shear tests. More importantly, this 3-dimensional surface representing shear strength of LHS-1 allows for the general prediction of shear strength at various combinations of normal stress and density for the material. Such a 3-dimensional representation, while new to the field of geotechnical measurements, is also powerful when comparing different granular materials, or when examining the effects of absorbed atmospheric water as discussed in Section 4.4 below.

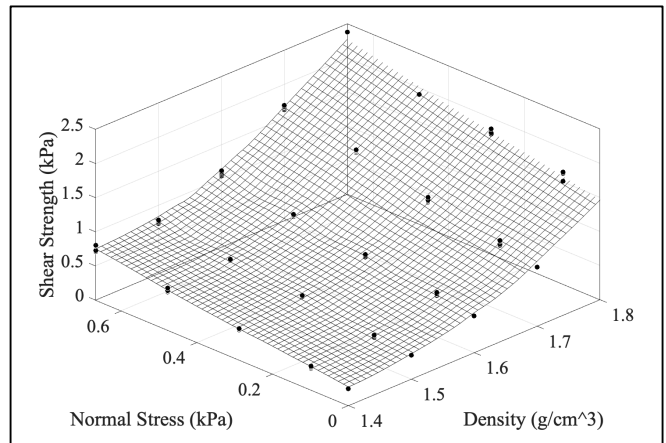
*Figure 8. Shear strength model for LHS-1 as a function of normal stress and density with direct shear test results.*

**4.0 Discussion**

The measured values of shear strength and cohesion are consistent with those observed during previous lunar and Martian space missions (Ming, 1992; Shoemaker et al., 1968). The results of this study demonstrate that density, particle size distribution, and surface coating such as moisture content can affect a sample's cohesion and shear strength. Through the analysis of empirical data, the generalized equations presented above allow for the estimation or prediction of cohesion and shear strength when using lunar and Martian regolith simulants, in terrestrial atmosphere and gravity, with normal stresses less than 0.68 kPa. While the shear strength and cohesion of a regolith simulant may change greatly based on density, particle size distribution, and absorbed water content, the results of this study are important for understanding and modeling PSI effects, as these same physical properties likely play an important role in viscous erosion and crater development. It is important to note that the shear strength and cohesion of granular regolith materials may behave differently when exposed to an environment with reduced gravity or vacuum conditions, as well as during exposure to electrostatic charge. When considering experiments or tests with lunar and Martian regolith simulants conducted on Earth, as a representation of actual lunar or Martian applications, special care should be taken to account for such conditions. Nevertheless, the results of this study not only address the modeling and simulation of PSI and ISRU tests on Earth, but also the preparation and simulation of surface activities for future space exploration missions using regolith simulants. Ultimately, the generalized equations for shear strength and cohesion of lunar and Martian regolith simulants presented above highlights the sensitivity of key parameters in geotechnical properties, including bulk density, particle size, particle shape, and absorbed atmospheric water as discussed in more detail below.

4.1 Effects of Density on Shear Strength and Cohesion

The results of this study demonstrate that bulk density has a large impact on both the shear strength and cohesion of granular materials, including lunar and Martian regolith simulants. Importantly, these results reveal that using a constant value for shear strength and cohesion is not appropriate when measuring or modeling PSI and ISRU applications, as an increase in bulk density causes an exponential increase for both of these critical material properties. These results show how increasing compaction, or increasing bulk density, can have an exponential impact on the strength and cohesion of a sample. Such results suggest that when building landing pads or structures on the Moon, compaction of regolith will greatly increase the strength and cohesiveness of the surface material.

With regard to direct shear tests, it is possible that when conducting shear measurements using normal stresses as small as 0.5 kPa, the subsequent compaction of regolith simulants will therefore yield a higher value for shear strength and cohesion than with smaller normal stresses (see Section 4.5 below). This effect of density on shear strength and cohesion is also likely responsible for the wide range of reported values for cohesion and shear strength from previous studies, which examined both Apollo regolith and lunar regolith simulants. Such studies have reported measured values of cohesion that range by more than 2-3 orders of magnitude, depending on the measurement technique and normal stress applied (Newson et al., 2021; Prabu et al., 2021). Therefore, when attempting to conduct direct shear strength measurements on granular materials, such as lunar and Martian regolith simulant, special care should be taken to choose an appropriate range of normal stresses commensurate with the intended mission application. Additionally, when reporting the shear strength and cohesion of a particular granular sample, it is therefore important to consider and report the bulk density at which the measurement was taken.

4.2 Effects of Particle Size on Cohesion and Shear Strength

As shown in previous studies, the particle size distributions of the examined Exolith and NASA SDL regolith simulants are consistent with measurements taken during lunar and Martian exploration missions (Easter et al., 2022; Kedar et al., 2017). It is important to note that particle size distributions were not intentionally varied for a given sample in this study, but rather a wide range of multi-sized particle materials were examined to establish fundamental, empirical relationships. While the results from previous studies suggest similarities between Exolith and NASA SDL regolith simulants, this study presents both traditional fit coefficients for particle size distributions, as well as a novel technique for characterizing particle sizes: the FWHM parameter. The results of this study also show that an increase in FWHM, meaning an increase in non-uniformity of particle size, corresponds to an increase in the rate of exponential growth of cohesion as a function of density. Given the relationship between cohesion and shear strength in Equation 1, this also implies that a broader range of particle sizes can also increase the shear strength of a granular material as a function of density as well.

Consistent with previous studies, having a more broadly distributed particle size distribution, including some amount of smaller grained particles less than roughly 50 μm, allows for larger ranges of achievable bulk density and porosity (Easter et al., 2022). This is directly supported by measurements of this study, such as the smaller range of density values that are achievable with glass beads (near-homogeneous, spherical particles) when compared to nonuniform or mixed samples such as regolith simulants. Since increased bulk density plays an important role in the cohesion and shear strength as discussed above, a wide mixture of particle sizes helps to amplify the effects of density on these key material properties. As such, when building on the Moon or Mars, a more diverse mix of particle sizes will therefore result in larger density effects on regolith cohesion (a higher exponential growth constant), increasing the shear strength and cohesiveness of the material even more with higher bulk densities.

While having a wide range of particle sizes can help increase the shear strength and cohesion of regolith structures, the inclusion of larger, coarse materials, such as agglutinates or pebble materials, may limit the ranges of achievable bulk densities, and therefore reduce the overall shear strength and cohesion of the sample. For example, as noted in Section 3.3, the addition of 25% coarse, anorthosite, agglutinate materials into LHS-1 reduced the range of achievable densities and subsequent material cohesion by roughly 10% for LHS-1-25A at high relative densities. Therefore, having a wide range of particles that is broadly distributed and well mixed to minimize pore space will increase the shear strength and cohesion of granular materials at higher bulk densities. Since the examined granular materials generally have particle sizes below 1mm, future studies should examine the effects of larger coarse materials (>1 mm) on the cohesion and shear strength of regolith simulants.

Also of note, the generalized equations above suggest that a measurable change in shear strength and cohesion, within the model errors discussed in Section 4.5 below, is possible with as little as roughly 15% change in FWHM. These results highlight the importance of choosing an appropriate particle size distribution when conducting experiments or simulations on Earth with regolith simulants where the shear strength and cohesion are important, such as during PSI tests, ISRU simulations, instrument design, fabrication of structures using regolith, traverse planning, or rover wheel design.

4.3 Effects of Particle Shape on Shear Strength and Cohesion

As discussed in Section 3.1, the measured particle shapes for both Exolith and NASA lunar regolith simulants are relatively consistent with actual lunar particle shapes (Tsuchiyama et al., 2022). In the context of Mars, while samples from the Martian surface have yet to be returned to Earth, erosion processes on Mars likely allow for similar particle shapes as represented by Exolith simulants. However, additional research is required to characterize the actual particle shapes on Mars and how they compare to regolith simulants. It is important to note that particle shapes were not intentionally varied for a given sample in this study but were directly measured across a series of different regolith simulants to examine any obvious relationships. Upon initial inspection, there do not appear to be any clear and measurable trends between examined particle shapes and the geotechnical properties of shear strength and cohesion.

While it is posited that more irregular and elongated particle shapes may be encountered on the Moon, potentially impacting the material's bulk density, shear strength, and cohesion, the particle shapes measured in this study compared to actual lunar samples suggest that these simulants still offer a reasonable representation of some bulk geotechnical properties. Yet, future studies should examine the impact of particle shapes on density, cohesion, and shear strength in more detail, with a wider range of particle shapes than examined in this study. Such future studies should also include the importance of elements such as nanophase iron or highly irregular particle shapes, that result from exposure to micrometeorites or space weathering effects. While it is believed that particle shape may have a relatively small impact on the shear strength and cohesion of a regolith simulant, it appears that using an appropriate density and particle size distribution is ultimately more crucial for replicating geotechnical properties of regolith simulants on Earth. Nevertheless, particle shapes presented in this study, including aspect ratio, elongation, and perimeter, likely play an important role in PSI events with lunar and Martian simulants on Earth, as these parameters will change the drag coefficient of particles being transported through an atmosphere.

4.4 Effects of Absorbed Atmospheric Water on Shear Strength and Cohesion

In order to determine the effects of absorbed atmospheric water on the shear strength and cohesion of regolith simulant, a 3-dimensional surface that represents the shear strength as a function of normal stress and bulk density (such as the one shown in Figure 8) was initially developed for LHS-1D, stored under atmospheric conditions. As described in Section 2.5, a similar 3-dimensional surface was then produced for LHS-1D after removing roughly 0.4 wt.% atmospheric water by mass when baking the sample. These two 3-dimensional surfaces, for dried LHS-1D and LHS-1D exposed to atmospheric water, were then subtracted from each other to show the changes in shear strength for LHS-1D as shown in Figure 9 below. It is important to note that the cohesion in Figure 9A is represented by the area of the plot where normal stress is equal to 0 kPa, which is also highlighted in Figure 9B. From these results, it is clear that atmospheric water causes non-monotonic changes in shear strength and cohesion, based on sample density and normal stress applied. For midrange densities, only ~0.4 wt.% water absorbed can increase cohesion by roughly 15%. However, for lower densities with high normal stress, the same amount of absorbed water has a smaller effect on shear strength. For normal stresses above roughly 0.4 kPa, it appears that the resultant shear strength of LHS-1D dominates when compared to any effects of absorbed atmospheric water.

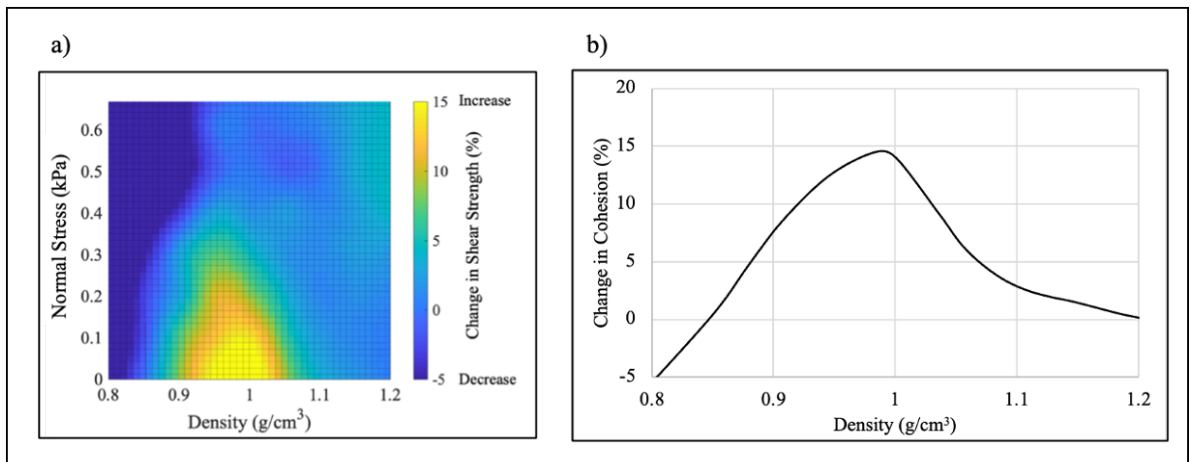

*Figure 9. Humidity effects on LHS-1D with ~0.4% absorbed, atmospheric water on:*
*a) shear strength and b) cohesion (normal stress of 0 kPa in contour plot on the left).*

As a potential explanation for this behavior, it is possible that water has a smaller effect on interparticle interactions at higher densities when compared to the mechanical interlocking of particles, which likely dominates cohesion with increased compaction. Conversely, for lower bulk densities, particles are far enough away that mechanical interlocking, as well as the effects from absorbed atmospheric water are both small. Thus, the largest impact from atmospheric water appears to occur near mid-ranged bulk densities where constituent particles are close enough to support absorbed water as well as interact with neighboring particles to ultimately increase shear strength of a sample.

While these limited shear tests with LHS-1D were used to demonstrate an initial understanding of the effects of absorbed atmospheric water on shear strength and cohesion, future studies should consider using a vacuum environment to further examine these effects. Additionally, since these limited tests involving removal of atmospheric water were only conducted using LHS-1D, it remains unclear if these non-monotonic effects or their magnitude are the same for granular samples with different particle size distributions. With larger particle sizes, or more uniform distributions of particle sizes, water may have a different ability to permeate into granular material and thus impact the interparticle forces in a sample differently. As such, future studies should examine the impacts of absorbed atmospheric water on other regolith simulants as well, including those with differing particle size distributions. Similarly, future studies should investigate the extent to which the amount of absorbed atmospheric water can impact the cohesion and shear strength of regolith simulants. While additional investigations are needed, these early results nevertheless highlight the importance of atmospheric water and humidity when considering regolith properties on Earth as a simulation for space missions.

4.5 Model Errors

While a rough linear fit has been used in Figure 7A to empirically fit the exponential growth constant, k, as a function of FWHM (producing Equation 5), this linear fit has a slope close to zero. When considering the confidence interval of the slope in Equation 5, the exponential growth constant could actually decrease with FWHM (linear fit with negative slope). The data and linear fit indicates that k may actually be independent of FWHM, or has a very weak increasing relationship. However, more data with greater precision is needed to establish a more definitive relationship between the parameter k and FWHM. Yet, in an effort to quantify the error the weak linear fit of exponential growth constant as a function of FWHM, a rough sensitivity analysis was conducted by substituting values into Equation 6 for LHS-1 (as derived from Equations 4 and 5). From these calculations, an increase in FWHM by roughly 10% results in an overall increase in cohesion and shear strength by roughly 2-3%. Thus, while the relationship between k and FWHM may not be perfectly linear, the contribution to the overall cohesion and shear strength model is small. Therefore, while more data with better precision is needed to definitively show the relationship between exponential growth constant and FWHM, this general approach and rough empirical fit provides a starting point for future studies.

The equations presented for shear strength and cohesion above are generally limited in their application to terrestrial atmosphere and gravity, with normal stresses less than 0.68 kPa, as these were explicitly measured during this study with empirical data. While these equations may demonstrate the inherent behavior of granular materials to higher normal stresses, bulk densities, and different particle size distributions, a careful consideration of error from such empirical measurements should be considered. In order to determine the error associated with the generalized equations for shear strength and cohesion presented above, standard error was plotted using partial derivatives of the shear strength equation presented in Equation 6, as well as the standard deviation of each associated parameter as shown in Equation 7, where n represents each general variable or fit parameter involved in Equations 2-6. From this approach, it is possible to quantify the standard error predicted by Equation 6, which varies based on the uncompressed bulk density, bulk density, normal stress, and fit parameters of interest.

$$\Delta \sigma_S^2 = \left(\frac{\partial f}{\partial \sigma_N}\right)^2 \Delta \sigma_N^2 + \left(\frac{\partial f}{\partial \varphi(\rho)}\right)^2 \Delta \varphi(\rho)^2 + \left(\frac{\partial f}{\partial \rho}\right)^2 \Delta \rho^2 + \ldots \left(\frac{\partial f}{\partial n}\right)^2 \Delta n^2 \qquad (7)$$

An example of this standard error is shown in Figure 10 for LHS-1 at different bulk densities and normal stresses. It is clear from this figure that these generalized equations have an error as large as 35% for higher normal stress and bulk densities, but generally decrease to less than a few percent for no normal stress applied (i.e., cohesion), as these were directly measured and calculated. As such, this highlights that while the presented equations provide a general approximation for shear strength of regolith simulants, the error of these equations increases with increased normal stresses. Such an increase in error, as well as the curved shape in Figure 10 below, is likely related to the uncertainties associated with internal friction angle measurements as a function of density.

It was also noted during direct shear measurements that larger normal stresses above 0.5 kPa inadvertently caused compaction of the sample material, increasing the bulk density and likely changing the resultant internal friction angle and cohesion. This effect was most notable for direct shear measurements with lower bulk densities (generally less than 1.40 g/cm$^3$), where the sample container could compress by as much as 2-3 mm with normal stresses of 0.50 to 0.68 kPa. For this particular experimental setup, this corresponds to a change in bulk density of roughly 0.04-0.06 g/cm$^3$, when accounting for the same amount of mass in a smaller volume. Using the generalized equations presented above, such a small increase in bulk density appears to change the resultant shear strength predictions by roughly 14%; generally, within the standard error of the model as shown in Figure 10 for normal stresses above 0.5 kPa. However, it is believed that larger normal stresses, especially with experimental setups that leverage hydraulic presses, would cause even larger changes in bulk density, and therefore resulting shear strength and cohesion. Future studies may improve on the predictive error of such generalized equations by examining the relationship between internal friction angle with bulk density and particle size in more detail. Nevertheless, while generalized equations for shear strength demonstrate increased errors with higher normal stresses, the presented equations for cohesion are well suited for predicting cohesion of generalized granular materials where only the uncompressed bulk density and particle size distribution are known.

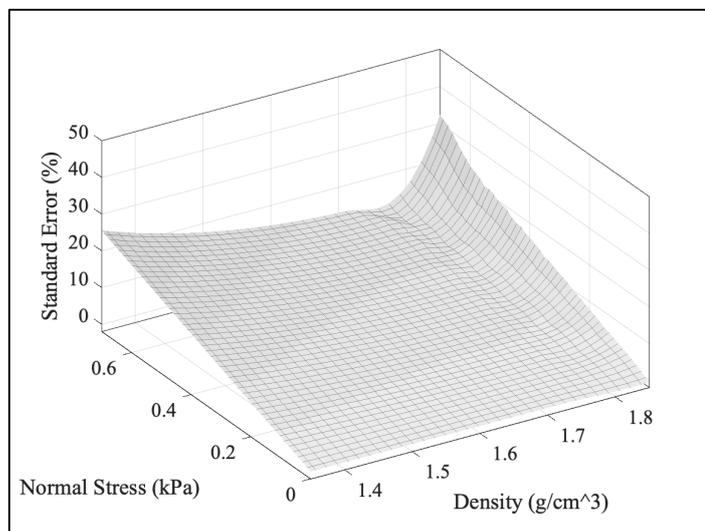

*Figure 10. Standard error for shear strength of LHS-1, as a function of density and normal stress.*

## 5.0 Conclusion

Generalized equations for the shear strength and cohesion of granular materials have been reported using empirical data from direct shear tests of applicable lunar and Martian regolith simulants. The reported generalized equations can be applied to granular materials where only the uncompressed bulk density and particle size distribution are parameters that need to be determined or estimated, potentially through remote sensing. The geotechnical properties of shear strength and cohesion for regolith can vary based on bulk density, particle size distribution, mineralogy, and absorbed atmospheric water content, and should be considered when attempting to use Earth-based measurements as a reference for space exploration. These generalized equations for shear strength and cohesion, as well as reported particle shapes and sizes, should be considered during PSI and ISRU applications. These results may also have implications during the planning and simulation of surface activities for space exploration missions. It is important to select a regolith simulant with the appropriate particle size distribution, mineralogy, and bulk density commensurate with intended mission applications, in order to accurately replicate geotechnical properties such as shear strength and cohesion. Future studies should further examine the effects of absorbed atmospheric water by repeating measurements under vacuum conditions, with different particle size distributions, and with different amounts of absorbed water. Additional investigations into the effects of larger coarse grains, as well as effects from irregular particle shapes, on geotechnical properties are also needed. Non-linear effects associated with cohesion should also be considered in future studies as well. Nevertheless, the generalized equations and information presented in this study allow for a better characterization of lunar and Martian regolith simulants, when attempting to simulate potential space missions on Earth.

**6.0 Declaration of Competing Interest**

The authors declare that they have no known competing financial interests or personal relationships that could have appeared to influence the work reported in this paper.

**7.0 Acknowledgments**

This work is supported by the Center for Lunar and Asteroid Surface Science (CLASS) under NASA cooperative agreement #80NSSC19M0214.